# Age Estimates of Universe: from Globular Clusters to Cosmological Models and Probes

Hira Fatima[1], Muzammil Mushtaq[2] and Syed Faisal Ur Rahman[3]
Institute of Space and Planetary Astrophysics (ISPA)
University of Karachi
Karachi, Pakistan
[1]hirafatima1407@gmail.com
[2]muzammilmushtaque@rocketmail.com

[3]faisalrahman36@hotmail.com

ABSTRACT

We performed the photometric analysis of M2 and M92 globular clusters in g and r bands of SLOAN photometric system. We transformed these g and r bands into BV bands of Johnson-Cousins photometric system and built the color magnitude diagram (CMD). We estimated the age, and metallicity of both the clusters, by fitting Padova isochrones of different age and metallicities onto the CMD.

We studied Einstein and de Sitter model, bench mark model, the cosmological parameters by WMAP and Planck surveys. Finally, we compared estimated age of globular clusters to the ages from the cosmological models and cosmological parameters values of WMAP and Planck surveys.

## 1. INTRODUCTION
### 1.1. Age of Universe from Globular Clusters

Stars are formed in clusters hence each star of the cluster has same age and chemical composition and at same distance from us. However the evolutionary lifetime of star depends on its mass. The two typical kind of star clusters are Globular and open clusters. Globular clusters found in galactic halo, they are distributed spherically around the galactic center, in the Sagittarius–Scorpius– Ophiuchus region (Fusi-Pecci & Clementini, 2000), while the open clusters are found in the plane, spiral arms and galactic nuclei.

Globular clusters are compact aggregations of about $10^5$-$10^7$ stars. Due to high stellar density (about few $10^3$ stars $ly^{-3}$) it is impossible to resolve the individual stars. Recently the state of art, the Hubble telescope allowed astronomers to dig into the very central regions of many globular clusters (Fusi-Pecci & Clementini, 2000). The diameter of these clusters are about 5-30 Parsecs. The spectroscopy of the globular clusters member stars show that they are deficient in heavy elements for instance carbon, oxygen and iron. This deficiency suggests that these are old star clusters (Population II) formed from primordial hydrogen and helium. The age estimates of globular clusters are about $10^{10}$ years (Phillips, 2013). Our Milky Way galaxy hosts about 200 globular clusters (Fusi-Pecci & Clementini, 2000).

Open clusters are loosely bond system of about $10^2$-$10^3$ stars. The diameter of open clusters ranges from 1-5 Parsecs. The studies of spectra of member stars suggests that they are rich in heavy elements such indicating that they are young stars (population I) with ages range from $10^6$-$10^9$ years (Phillips, 2013).

The first 'astronomical' detection of globular clusters is credited to John Herschel. In the 1830s, Herschel realized that a large number of these clusters are concentrated in a relatively small region of the sky towards Sagittarius. Later on, Harlow Shapley detected variable stars in several globular clusters. In 1917, Shapley got the idea that the Galactic center is very distant from the Sun and it is in the direction of Sagittarius. He estimated the size of the Milky Way. Edwin Hubble pioneered the search for globular clusters in the local group of galaxies. He detected about 100 Globular Clusters in the Andromeda galaxy however, recently more than 350 globular clusters are known (Fusi-Pecci & Clementini, 2000).

The universe began in a Big Bang about 13.7 years ago. At that time, the matter in the universe only consists of Hydrogen and Helium, produced by big bang nucleosynthesis. Hence the first stars formed had no metal content and this metal content increased with successive star generations. Based on this metal content, the stars formed immediately after the Big Bang are known as Population III stars, while metal poor stars are known as population II, and metal-rich stars are known as population I stars (Carroll & Ostlie, 2006).

Globular clusters are among the oldest observable objects in the universe considered as the remnants of galaxy formation. They are used to find the galactic center (Harris, 1976; Reid, 1993) and age of universe (Krauss & Chaboyer, 2003). The ages of these old clusters can be compared and contrast to the estimated ages of universe from Hubble constant and other cosmological models (Chaboyer, 1994; Krauss & Chaboyer, 2001; Salaris, Degl'Innocenti, & Weiss, 1997).

We selected M2 and M92 globular clusters for photometric analysis. Globular cluster M2 (NGC 7089) in Aquarius, is a class II cluster. It was first discovered by Maraldi in 1746. After 14 years, Charles Messier independently rediscovered and cataloged it as a "nebula without stars." Finally William Herschel was the first to resolve it into stars.

M92 in Hercules is a class IV cluster, it is one of the original discoveries of Johann Elert Bode, who found it in 1777. In 1981, Charles Messier independently rediscovered it and cataloged it. It was William Herschel who first resolved it into stars in 1783. More information about these globular cluster can be found on Messier SEDS website (Frommert, 2014).

For cluster age determination the first step is photometric analysis. In aperture photometry technique, there are two radii. The first smallest radius is centered on the star know as radius of measurement. The next radius are the inner and outer radius of the sky annulus. Aperture photometry calculates the flux of each star by summing up the pixel values within the measurement radius, then measure the sky signal per pixel (in average, mode or median) within the annulus radius (see figure 3). Then, the flux is converted in instrumental magnitude by applying the following equation.

$$m_{ins} = -2.5 \times \log \frac{(I_* - n_{px} \times I_{sky})}{t_{exp}} \qquad (1)$$

Where, $I_*$ is the sum of counts within the measurement radius, $n_{px}$ is the number of pixels contained in that circle, $I_{sky}$ is the average intensity of the sky per pixel calculated within the wider ring, and $t_{exp}$ is the exposure time for the image (in seconds) (Barolo, Bo, & Naibo, 2011; Romanishin, 2002).

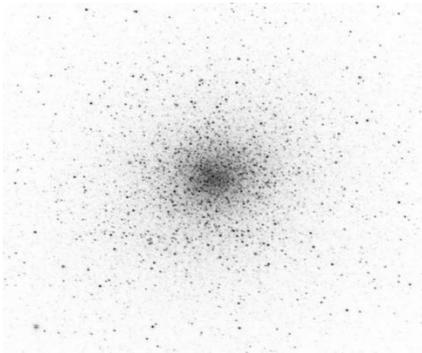

Figure 1 *M92: Optical SDSS DR7 g band*

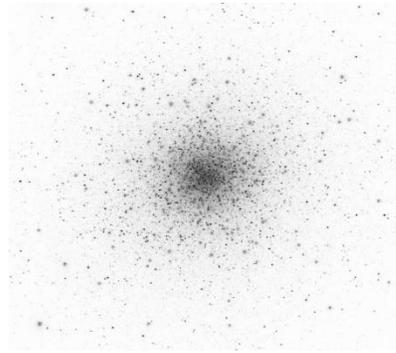

Figure 2 *M2: Optical SDSS DR7 g band*

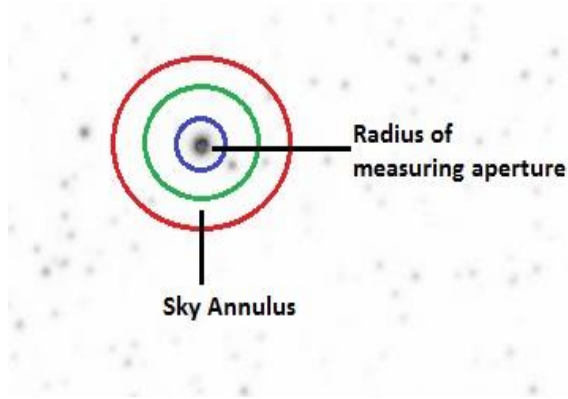

Figure 3 *Aperture Photometry Illustration*

In crowded fields like globular clusters, there is not enough empty spaces around the stars or in other words there would be many stars in the sky annulus so it would not be possible to get a good sky value using this technique. In crowded fields, where stars are too close or overlapped, the PSF (Point Spread Function) photometry, allowed us to discriminate one star from another. To do so, a mathematical model of the PSF is built based on the selected stars of our cluster. The PSF model basically represents the light distribution of a point-like source modified by atmospheric effects (Barolo et al., 2011).

Hertzsprung-Russel diagram, illustrates the correlation between the two main observational properties of stars i.e. luminosity and surface temperature. In this two dimensional plot, luminosity is shown on vertical axis while the surface temperature on horizontal axis. Mostly in H-R diagrams luminosity of star is represented by its magnitude and its surface temperature by its spectral type (Phillips, 2013). In particular, if the surface temperature is expressed in radiation difference measured from two different wavelength bands (say B-V color index), such a diagram is referred to as a color-magnitude diagram (CMD). CMD help us link the observational and theoretical calculations of stellar evolution and calculate the significant parameters: age, metallicity, reddening and distance of the cluster.

An isochrone is a mathematical model representing an HR diagram at fixed age and metallicity. The Padova database of evolutionary tracks and isochrones is a repository of published tracks, isochrones, bolometric corrections, chemical yields, etc. However some new databases are also available such as Cstars, dustyAGB07, and YZVAR(Girardi, Grebel, Odenkirchen, & Chiosi, 2004). The most important point for the age determination is the overlapping of different isochrones on observed diagram to find the one that best fitted in particular turn-off region. Further the metallicity, reddening and distance can also be determined from this isochrones fitting.

## 1.2. Age of Universe from Cosmological Models and Sky Surveys

In 1929, Edwin Hubble through his observations shows that greater the distances, greater the apparent speeds of the recession galaxies, that concluded the universe is expanding uniformly in every direction. So, it is possible to extrapolate back in time till the universe had no size i.e., its origin, to estimate the age of the universe, this is known as Hubble age of the universe and it can be written as $\left(t_H = \frac{1}{H_0}\right)$ (Morison, 2013).

However, in 1922 Alexander A. Friedmann used Einstein's field equations to primarily describe the contraction or expansion of the spatially homogeneous and isotropic universe as a function of time. Friedmann equation in the form of time is written as:

$$t_u = \frac{1}{H_0} \int_0^a \frac{da}{\left[\frac{\Omega_{r,0}}{a^2} + \frac{\Omega_{m,0}}{a} + \Omega_{cc,0} a^2 + (1-\Omega_0)\right]^{\frac{1}{2}}} \qquad (2)$$

Here, the $t_u$ is the age of universe, $H_0$ is current value of Hubble parameter called Hubble constant, $\Omega_0$ is the current total density parameter has value greater, less or equal to 1 which describes the curvature of universe (*k*). In reality, the evolution of universe is complicated by the fact that it contains different dominated energy density components (radiation, matter, cosmological constant) in different eras which played a big role in the evolution of the universe those are, $\Omega_{r,0}, \Omega_{m,0}$ and $\Omega_{cc,0}$ which corresponds to the current density parameters of radiation, matter (including baryonic and non-baryonic) and cosmological constant respectively. The integral limit is from 0 to $a$ where, $a$ is the scale factor has the relationship with redshift i.e. $\left(a = \frac{1}{1+z}\right)$ (Ryden, 2003).

We calculated the age of universe analytically by using different cosmological models and observations and compared it with the age of oldest globular clusters M2 and M92 situated in Milky Way galaxy. The significance of this work is that the age of the globular clusters estimate the lower bound of the age of universe, better understanding the stellar evolution, evaluate the difference between the analytical and observational results.

## 2. OBSERVATIONAL DATA

This section gives a brief background of two main objectives of this report i.e. the age determination of universe from globular clusters, sky surveys and cosmological models.

### 2.1. Globular Cluster Age Estimation

The Sloan Digital Sky Survey mapped one quarter of the entire sky and performed a redshift survey of galaxies, quasars and stars. Data Release 7 (DR7) is the seventh major data release and provides images, imaging catalogs, spectra, and redshifts for download (Abazajian et al., 2009). The data used in this project was Sloan Digital Sky Survey, data release 7 images, acquired from NASA SkyView website (NASA SkyView, 2016) and summarized in table 1 and 2. We worked on g and r bands of ugriz photometric system, so high resolution (1024 pixels) images of only these bands were acquired.

CMD is an interactive web service which provides interpolated isochrones for any age and metallicity and the photometry data can be produced for many different band systems (Girardi, 2016). We did thorough literature review of photometric analysis of globular clusters under study (Carney, Storm, Trammell, & Jones, 1992; Lee & Carney, 1999; Salaris et al., 1997; Stetson & Harris, 1988) and then downloaded Padova isochrones with suitable age and low metallicities in UBVRIJHK photometric system and Chabier (2001) lognormal initial mass function.

For magnitude calibration and creation of color magnitude diagram, the photometric parameter values of images are needed. We worked for photometric values of the images used in this project. By using the coordinates of under study globular clusters, the field calibration and statistics data was obtained from SDDS Data Archive Server (SDSS, 2013), which was processed in TOPCAT to get the values of desired parameters (atmospheric extension coefficient, air mass and photometric zero point). The parameter values of images used in this project are given in table 2.

*Table 1 M2 and M92 globular clusters profile*

| Serial # | Globular Cluster | | Right Ascension | Declination | Class | Constellation |
|---|---|---|---|---|---|---|
| | Messier Name | NGC Name | RA | Dec | | |
| 1 | M2 | NGC7089 | 21 33 27.0 | -00 49 23 | II | Aquarius |
| 4 | M92 | NGC6341 | 17 17 7.3 | +43 08 09 | IV | Hercules |

*Table 2 Summary of SDSS DR7 data used in this project*

| Serial # | Globular Cluster Messier Name | Band | Air mass | Atmospheric absorption coefficient | Photometric Zero point | Exposure time | Pixel size | Data Source |
|---|---|---|---|---|---|---|---|---|
| | | | | k | $m_0$ | Sec | arcsec | |
| 1 | M2 | g | 1.023 | 0.163 | 24.412 | 53.9 | 0.396" | Sloan Digital Sky Survey Data Release 7 |
| | | r | 1.019 | 0.087 | 23.028 | | | |
| 4 | M92 | g | 1.039 | 0.163 | 24.432 | | | |
| | | r | 1.038 | 0.087 | 23.005 | | | |

## 2.2. For Age of Universe from Sky surveys and Cosmological Models

The value of cosmological parameters from sky surveys were taken from the published results of these surveys (Adam et al., 2015; Hinshaw et al., 2013)

## 3. WORK DESCRIPTION

In this project we used several softwares such as IRAF (Image Reduction and Analysis Facility) for photometric analysis of images, DS9 to visualize images, TOPCAT to manipulate data tables, MS Excel for calculations, and MTALAB for plotting color magnitude diagram and fitting isochrones and Python for solving Friedmann equation.

## 3.1. Photometric determination of age and metallicity of M92 and M2 globular clusters

The methodology adopted here have been used in several photometric studies of open and globular star clusters (Barolo et al., 2011; Elena Boldrin, 2009; Francesco Bussalo, 2012; Francesso Battaglini, 2011). In this section a brief account of methodological framework is given, which was applied on g and r band images of M2 and M92 globular cluster.

First, the instrumental magnitude of stars were obtained using aperture photometry technique (discussed in introduction section). We used imexam command in IRAF and Ds9 to examine 15 stars spread across the image (in order to obtain the average Full Width at Half Maximum (FWHM) of the images) and 15 background (sky) points to get the average value of sigma.

The obtained average FWHM value was used to execute the DAOFIND command in IRAF, this task gave the coordinates of recognized stars. We used TVMARK command in IRAF, to overlay the recognized star from coordinate file on the image.

To calculate the instrumental magnitude of all stars based on aperture photometry technique, we used PHOT command in IRAF. This gave a text file which contains the coordinates of every star with its corresponding magnitude.

Second scan of image was done using PSF photometry. Using the PSTSELECT task in IRAF, about 25 isolated stars were selected. Then, using the PSF task, the PSF model was built by selecting only the best stars with the similar surface and radial plots. The PSF model was applied on the image using ALLSTAR task in IRAF, this scaned the image for given no. of iterations, rejected merged stars, selected isolated stars and calculated their instrumental magnitudes. Finally, we obtained a text file with the coordinates of selected stars and their corresponding instrumental magnitudes.

The same procedure was applied on images of r and g filter of both globular clusters. To detect the common light sources in both r and g bands/filters, the stars detected in two bands were compared by using TOPCAT software. 2D Cartesian algorithm in Pair match task was used to get the text file of instrumental magnitude of common light sources in the two bands. The instrumental magnitudes were transformed into the calibrated ones by applying the following equations in MS Excel:

$$m_{cal} = m_0 + m - k_m \times airmass \qquad (3)$$

Where, $m_{cal}$ is calibrated magnitude, $m_0$ is photometric zero point, m is the instrumental magnitude and $k_m$ is the atmospheric absorption coefficient (See table 2). We then performed conversion from the ugriz to the UBVRI photometric system (by Johnson Cousins) using these formulas in MS Excel

$$B = g + 0.349 \times (g - r) + 0.245 \qquad (4)$$

$$V = g - 0.569 \times (g - r) + 0.021 \qquad (5)$$

We used these data to build the CMD in MATLAB. We overlapped different isochrones on our observed diagram to find the one that best fitted in particular turn-off region, which is the most important task in age determination.

## 3.2. Determination of Age Universe from Sky Surveys and Cosmological Models

Many cosmological models have been developed to better understand the evolution of universe and parameters it contains. One of the big tasks for cosmologist is to find the accurate age of universe, for which the density parameters and expansion rate of universe must be initially known. Therefore, theoretical assumptions were made for the values of density parameters to estimate the behavior and fate of assume universe. In our worked, we used Einstein De-sitter Universe model, Benchmark model, and Lambda CDM model to calculate the age of universe. On the other hand, Observations of cosmological parameters done by WMAP and Planck Satellites play a key role to better understand the evolution of universe which also improved the theoretical models. We solved the Friedmann equation for time using Python language. The scipy.integrate sub-package provides several integration techniques in python and quad is used for general purpose integration where a function contains one variable between two points. The python code used in this paper is given in annexure section.

## 4. RESULTS
## 4.1. Estimated Age of Universe using Globular cluster

About 8000 and 6000 common light sources (stars) were detected in both (g and r) bands of M2 and M92 respectively. The color magnitude diagrams of M2 and M92 were built using aperture and PSF photometry. Our analysis showed that within our photometric errors and the limitations of the Padova isochrones, an isochrones of age of 13.5 Gyrs and a metallicity Z=0.0005 was found to be the best fit the CMD of M92 Globular cluster. However the isochrones of age 13 Gyrs and metallicity Z= 0.0006 was found to be best fit to the CMD of M2 globular cluster.

Hence we can conclude that our values are in good agreement with different values from the literature. We are also working on M13 and M15 Globular clusters.

*Table 3 Summary of estimated significant parameters of M2 aand M92 globular clusters*

| Serial # | Globular Cluster | Estimated Age | Metallicity |
|---|---|---|---|
| | | Gyrs | Z |

| 1 | M2 | 13 | 0.0001 |
|---|----|----|--------|
| 2 | M92 | 13.5 | 0.0006 |

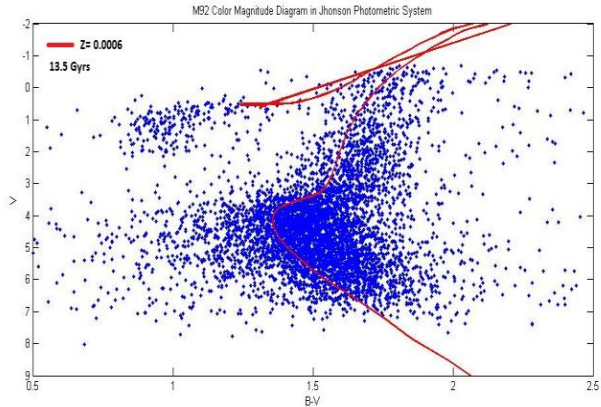

*Figure 4 M92 Isochrone fitting on CMD*

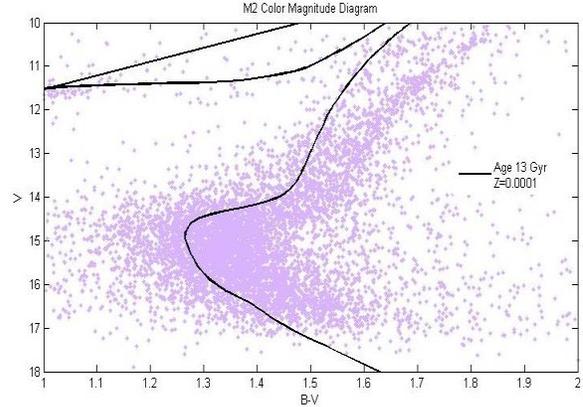

*Figure 5 M2 Isochrone fitting on CMD*

## 4.2. Estimated age of universe using cosmological models

A cosmological model represents the Universe at a particular scale. The ultimate aim of observations and theoretical models are to obtain the physical description of the Universe and to better understand the evolution of universe and the parameters it contains.

In 1932, Einstein and de Sitter derived a homogeneous and isotropic model in which they assumed a zero spatial curvature of the universe (flat universe) that contained non-relativistic matter only. Therefore, $\Omega_0 = \Omega_{m,0} = 1$ and $\Omega_{r,0} = \Omega_{cc,0} = 0$; for $H_0 = 67$ km/s/Mpc, then age of universe ($t_u$) is 9.73048 Giga yrs. Calculate age of universe from this model has even much lower value than the estimated ages of globular clusters. Therefore, curvature in universe should be considered for reliable estimate of the age. For $k = -1$, $\Omega_0$ should be less than 1, so, expansion will continue for all time, universe ends in the "Big Chill". However, negative curvature matter only universe could have the ages depend on $\Omega_0$;

*Table 4 Ages of the Universe at values of Ω_0 less than 1 in the Curvature Matter Only Universe.*

| $\Omega_0$ | $t_u$ (Gyrs) |
|---|---|
| 0.9 | 9.9339 |
| 0.5 | 10.9984 |
| 0.3 | 11.8047 |

For different $\Omega_0$ values the age of universe still do not reached at age of globular cluster. Thus, we can concluded that in present era, the effect of matter (both baryonic and non-baryonic) and radiations on the evolution of universe is least and cannot satisfy the bigger ages of globular clusters. Therefore, there must be some form energy latent within space itself that is totally homogeneous throughout space – called as the dark energy. A positive lambda term can be referred as a fixed positive energy density that expands all space and is unchanging with time. Therefore, the consequence of dark energy should become further noticeable as the universe ages and its size increases. The accelerated expansion of the universe discovery through type 1a supernovae surveys (Riess, 1998; Purlmutter 1999) hinted towards the presence of dark energy. Later cosmic microwave background surveys and discovery of phenomenon such as the integrated Sachs-Wolfe effect (Barreiro, 2013; Rahman, 2014; Hinshaw, 2013) further strengthened the validity of the presence of dark energy. Accelerated expansion particularly dominates the redshifts~1.5-2 (Giannantonio. 2008; Rahman. 2014). The cosmological constant is the simplest solution to the problems of cosmological acceleration and the age issue of universe. Lambda become a vital component in the current standard model of cosmology known as lambda-cold dark matter model that integrates both cold dark matter and the cosmological constant, it can be used to predict the evolution and the age of universe (Morison, 2013).

Benchmark Model is one of the best-fitted models on the observational data which contains both matter and cosmological constant. By taking current density parameter for matter i.e., $\Omega_{m,0} = 0.3$, then at $\Omega_{cc,0} = +0.7$; the universe is spatially flat and is destined to end in an exponentially expanding Big Chill. The age of the universe estimated from Benchmark Model is 13.4684 Giga yrs at $H_0$=70 Km/s/Mpc, that is much nearer to the ages of globular clusters and this enforced the acceptance of the domination of cosmological constant in Universe.

### 4.3. Estimated age of Universe from sky surveys/Satellites

Cosmological observations by WMAP and Planck's Satellite show the results;

*Table 5 Observed cosmological parameters are taken from WMAP (Seven and Nine year release data) and Planck's (Ade et al., 2014; Hinshaw et al., 2013; Weiland et al., 2010)*

| Observations | H0 (Km/s/Mpc) | $\Omega_{cc,0}$ | $\Omega_{m,0}$ | Age of Universe (Giga yrs) |
|---|---|---|---|---|
| WMAP (7 year data) | 70.4 | 0.728 | 0.2726 | 13.75 |
| WMAP (9 year data) | 69.32 | 0.7135 | 0.28648 | 13.772 |
| Planck's 2013 result | 67.80 | 0.692 | 0.315 | 13.798 |
| Planck's 2015 result | 67.74 | 0.6911 | 0.308 | 13.799 |

Observations through different satellites proved that cosmological constant has now a dominated role in the evolution of universe and largely effect the age of universe, which also satisfies the bigger ages of globular clusters that might be the lower bound of the age of the Universe.

## 5. CONCLUSION

The results described in this paper imply that dark energy density, is required to dominate the energy density of the universe, which plays a big role in increasing the age of the universe and justifying the age of globular clusters. Although, all fundamental independent observables in the modern cosmology including measurements of standard candles, large-scale structures, CMB measurements, and measured the age of universe – all are agreeing with a dark energy dominated cosmological model, involving a flat universe with about 30% of its baryonic and non-baryonic matter contents and about 70% dark energy.

ACKNOWLEDGEMENT

Mr. Ghulam Laghari, Section head Astrophysics, Pakistan Space and Upper Atmosphere Research Commission (SUPARCO), for giving us internship opportunity at SUPARCO and for his cordial support, valuable information and guidance, which helped us in completing this project.

ANNEXURES

Python code used for age estimation

```python
>>>from scipy.integrate import quad
>>>omega_m = 0.30     # Density parameter of Baryonic and Non-baryonic matter
>>>omega_r = 0.0    # Density parameter of Radiations and Relativistic particles
>>>omega_DE = 0.70     # Density parameter of Cosmological constant
>>>omega_o = omega_m + omega_r + omega_DE        # Total density parameter
>>>Ho = 67.0     # Hubble constant
>>>def integrand(a):
>>>    return 1/((omega_r/a**2 + omega_m/a + omega_DE*a**2 + (1-omega_o))**0.5)

>>>ans, err = quad(integrand, 7.2411e-4, 1)         # Lower limit of scale factor (a) is
                                                    # 7.2411e-4 when redshift (z) is 1380
                                                    # Upper limit of (a) is 1 when (z)
                                                    # is zero
# Convert the Hubble constant into time (years) and add ~300 kilo years because
# the time required from Big Bang to formation of CMB...
>>>print (ans*(1/(Ho/3.086e19))) / (60*60*24*365.25)+ 380000
```